\input tables.tex

\documentstyle[12pt]{article}
\def\Journal#1#2#3#4{{#1} {\bf #2}, #3 (#4)} 
 
\def\NPB{{\em Nucl. Phys.} B} 
\def\PLB{{\em Phys. Lett.}  B} 
\def\PRL{\em Phys. Rev. Lett.}

\catcode`\@=11
\long\def\@makefntext#1{
\protect\noindent \hbox to 3.2pt {\hskip-.9pt
$^{{\ninerm\@thefnmark}}$\hfil}#1\hfill}	

\def\@makefnmark{\hbox to 0pt{$^{\@thefnmark}$\hss}} 

\def\ps@myheadings{\let\@mkboth\@gobbletwo \def\@oddhead{\hbox{}
\rightmark\hfil\ninerm\thepage}
\def\@oddfoot{}\def\@evenhead{\ninerm\thepage\hfil
\leftmark\hbox{}}\def\@evenfoot{}
\def\sectionmark##1{}\def\subsectionmark##1{}}

\renewcommand{\thefootnote}{\fnsymbol{footnote}}

\newcounter{sectionc}\newcounter{subsectionc}\newcounter{subsubsectionc}
\renewcommand{\section}[1] {\vspace*{0.6cm}\addtocounter{sectionc}{1}
\setcounter{subsectionc}{0}\setcounter{subsubsectionc}{0}\noindent
{\normalsize\bf\thesectionc. #1}\par\vspace*{0.4cm}}
\renewcommand{\subsection}[1] {\vspace*{0.6cm}
\addtocounter{subsectionc}{1}
\setcounter{subsubsectionc}{0}\noindent
{\normalsize\it\thesectionc.\thesubsectionc. #1}\par\vspace*{0.4cm}}
\renewcommand{\subsubsection}[1]
{\vspace*{0.6cm}\addtocounter{subsubsectionc}{1}
\noindent {\normalsize\rm\thesectionc.\thesubsectionc.\thesubsubsectionc.
#1}\par\vspace*{0.4cm}}

\newcounter{appendixc}
\newcounter{subappendixc}[appendixc]
\newcounter{subsubappendixc}[subappendixc]

\renewcommand{\appendix}[1] {\vspace*{0.6cm}
\refstepcounter{appendixc}
\setcounter{figure}{0}
\setcounter{table}{0}
\setcounter{equation}{0}
\renewcommand{\thefigure}{\Alph{appendixc}.\arabic{figure}}
\renewcommand{\thetable}{\Alph{appendixc}.\arabic{table}}
\renewcommand{\theappendixc}{\Alph{appendixc}}
\renewcommand{\theequation}{\Alph{appendixc}.\arabic{equation}}
\noindent{\bf Appendix \theappendixc #1}\par\vspace*{0.4cm}}

\def\abstracts#1{{
\centering{\begin{minipage}{12.2truecm}\footnotesize\baselineskip=12pt\noindent
\centerline{\footnotesize ABSTRACT}\vspace*{0.3cm} \parindent=0pt #1
\end{minipage}}\par}}

\renewenvironment{thebibliography}[1] {\begin{list}{\arabic{enumi}.}
{\usecounter{enumi}\setlength{\parsep}{0pt} 
\setlength{\leftmargin 1.25cm}{\rightmargin 0pt}
\setlength{\itemsep}{0pt} \settowidth
{\labelwidth}{#1.}\sloppy}}{\end{list}}

\newcounter{itemlistc}
\newcounter{romanlistc}
\newcounter{alphlistc}
\newcounter{arabiclistc}

\newcommand{\fcaption}[1]{
\refstepcounter{figure}
\setbox\@tempboxa = \hbox{\footnotesize Fig.~\thefigure. #1} \ifdim
\wd\@tempboxa > 6in
{\begin{center}
\parbox{6in}{\footnotesize\baselineskip=12pt Fig.~\thefigure. #1}
\end{center}}
\else
{\begin{center}
{\footnotesize Fig.~\thefigure. #1}
\end{center}}
\fi}

\newcommand{\tcaption}[1]{
\refstepcounter{table}
\setbox\@tempboxa = \hbox{\footnotesize Table~\thetable. #1} \ifdim
\wd\@tempboxa > 6in
{\begin{center}
\parbox{6in}{\footnotesize\baselineskip=12pt Table~\thetable. #1}
\end{center}}
\else
{\begin{center}
{\footnotesize Table~\thetable. #1}
\end{center}}
\fi}

\def\@citex[#1]#2{\if@filesw\immediate\write\@auxout
{\string\citation{#2}}\fi
\def\@citea{}\@cite{\@for\@citeb:=#2\do
{\@citea\def\@citea{,}\@ifundefined
{b@\@citeb}{{\bf ?}\@warning
{Citation `\@citeb' on page \thepage \space undefined}} {\csname
b@\@citeb\endcsname}}}{#1}}

\newif\if@cghi
\def\cite{\@cghitrue\@ifnextchar [{\@tempswatrue
\@citex}{\@tempswafalse\@citex[]}}
\def\citelow{\@cghifalse\@ifnextchar [{\@tempswatrue
\@citex}{\@tempswafalse\@citex[]}}
\def\@cite#1#2{{$\null^{#1}$\if@tempswa\typeout
{IJCGA warning: optional citation argument ignored: `#2'} \fi}}

 1
 1
 1

\font\ninerm=cmr9

\def\be{\begin{equation}}
\def\ee{\end{equation}}
\def\bea{\begin{eqnarray}}
\def\eea{\end{eqnarray}}
\begin{document}
\def\FI{Fayet-Iliopoulos}
\def\ev#1{\langle#1\rangle}

\def\Fund#1#2{\vcenter{\vbox{\drawbox{#1}{#2}}}}
\def\Asym#1#2{\vcenter{\vbox{\drawbox{#1}{#2}
              \kern-#2pt       
              \drawbox{#1}{#2}}}}
 
\def\fund{\Fund{6.5}{0.4}}
\def\asym{\Asym{6.5}{0.4}}
\batchmode
  \font\bbbfont=msbm10
\errorstopmode
\newif\ifamsf\amsftrue
\ifx\bbbfont\nullfont
  \amsffalse
\fi
\ifamsf
\def\IR{\hbox{\bbbfont R}}
\def\IC{\hbox{\bbbfont C}}
\def\ID{\hbox{\bbbfont D}}
\def\IZ{\hbox{\bbbfont Z}}
\def\IF{\hbox{\bbbfont F}}
\def\IP{\hbox{\bbbfont P}}
\else
\def\IR{\relax{\rm I\kern-.18em R}}
\def\IZ{\relax\ifmmode\hbox{Z\kern-.4em Z}\else{Z\kern-.4em Z}\fi}
\def\IF{\relax{\rm I\kern-.18em F}}
\def\IP{\relax{\rm I\kern-.18em P}}
\fi
\def\f#1#2{\textstyle{#1\over #2}}
\def\half{\f{1}{2}}
\def\Tr{{\rm Tr}}
\rightline{IASSNS-HEP-97/79}
\baselineskip=17pt
\centerline{\normalsize\bf RENORMALIZATION GROUP FIXED POINTS}
\baselineskip=17pt
\centerline{\normalsize\bf AND CONNECTIONS WITH DUALITY}
\baselineskip=17pt
\centerline{\normalsize\bf IN VARIOUS DIMENSIONS}

\centerline{\footnotesize KENNETH INTRILIGATOR} \baselineskip=13pt
\centerline{\footnotesize\it Institute for Advanced Study}
\baselineskip=12pt
\centerline{\footnotesize\it Princeton, NJ 08540, USA}

\vspace*{0.9cm}
\abstracts{We review some recent results concerning gauge theories in
various dimensions.  In particular, we discuss RG fixed points and
``mirror'' symmetry duality in 3d $N=4$ supersymmetric gauge theories
and a classification of non-trivial RG fixed points in 5d $N=1$
supersymmetric theories.}

\vspace*{0.6cm}
\normalsize\baselineskip=15pt
\setcounter{footnote}{0}
\renewcommand{\thefootnote}{\alph{footnote}}

\section{Introduction}

By using exact results which can be obtained for supersymmetric gauge
theories, it has recently been seen that these theories exhibit a
variety of interesting phenomena and phases.  For example, non-trivial
RG fixed points arise in many susy gauge theories.  Also some exhibit
duality, i.e. universality: different UV theories can flow to the same
RG fixed point in the IR.  There is also, in some examples, the
phenomenon of composite gauge invariance: strong coupling dynamics can
lead to new gauge fields via collective excitations.  The classic
examples of these phenomena, due to Seiberg \cite{sem}, are $N=1$
supersymmetric QCD for various numbers of flavors and colors. These
theories have non-trivial RG fixed points for ${3\over
2}N_c<N_f<3N_c$.  There are dual $SU(N_f-N_c)$ gauge theories which
flow to the same RG fixed points.  There is composite, IR free,
$SU(N_f-N_c)$ gauge invariance for $N_c+2\leq N_f\leq {3\over 2}N_c$.
Other examples can be found, for example, in \cite{isrev} and
references therein.

As an aside, we point out that the composite gauge invariance does not
violate the theorems of Weinberg and Witten \cite{weiwit}, which
state: (1) Given a Lorentz-covariant, conserved $J_\mu$, there can be
no composite or elementary massless particles carrying its charge with
spin $j>{1\over 2}$; (2) If there is a Lorentz-covariant, conserved
$T_{\mu \nu}$, there can be no composite or elementary massless
particles with $j>1$.  The point is that the new massless composite
$j=1$ gauge fields do not carry charge under any Lorentz-covariant
conserved $J_\mu$.  The new gauge symmetry appears ``out of thin
air'' \cite{qEW}.

There are still many open questions, for example: Properties of RG
fixed points are not well understood in general.  Is there a better
way to describe fixed points than in terms of the original UV
Lagrangians?  What is the general statement of duality? What is the
duality operator map? Is there a framework where duality is manifest?
Is there a framework where duality is exact, even away from the fixed
point?

There has also been recent interest in exploring gauge theories in other
dimensions, both with the hope of gaining insight into these issues
and because of powerful new connections between gauge theories and
string theory.

Note that, in dimensions other than four, the gauge coupling is
dimensionful:
$$d^dx~{1\over g^{2}}~F_{\mu \nu}^2\sim {1\over g^{2}}~L^{d-4}, \quad
\rm{so} \qquad g\sim L^{(d-4)/2}.$$ Therefore, for $d<4$, {\bf all}
gauge theories are asymptotically free in the UV and strongly coupled
in the IR.  For $d>4$, {\bf all} gauge theories (with finite $g$) are
infrared free and have a ``Landau-pole'' singularity in the
ultraviolet.  They can only be regarded as the low energy limit of
some other theory.

While these statements are very general, different theories do exhibit
a wide variety of phenomena.  For example many theories, both in $d<4$
{\bf and} in $d>4$ have non-trivial RG fixed points.  Non-trivial RG
fixed points for $d>4$ mean that these gauge theories really
``exist.''  The IR free theory with finite $g$ can be regarded as a
perturbation away from the RG fixed point by ${1\over g^2}~F_{\mu
\nu}^2$.

\section{Composite Gauge Invariance in String Theory}

In string theory, singular geometry or gauge fields can lead to
composite gauge invariance or RG fixed points.  There are two
(related) basic ways in which non-perturbative enhanced gauge symmetry
arises in string theory.  One is via singular geometry or gauge fields
associated with compactification.  The other is via D-branes
\cite{Polrev}, which are collective excitations of various dimensions
with supersymmetric gauge theories living in their world-volume.

For type II strings, the supersymmetric gauge theories living in the
world-volume of D-branes have 16 super-charges.  For example, $N$
coincident type IIB 5-branes have a 6d ${\cal N}=(1,1)$ supersymmetric
theory living in their world-volume with gauge group $U(N)$.  The 6d
${\cal N}=(1,1)$ gauge multiplet consists of a ${\cal N}=(1,0)$ gauge
multiplet, which contains no scalars, and a ${\cal N}=(1,0)$ adjoint
matter multiplet.  The expectation values of the $4N$ scalars in the
Cartan of the adjoint matter multiplet, which break $U(N) \rightarrow
U(1)^N$, correspond to the locations of the $N$ 5-branes in the four
transverse directions.  The separation of branes gives a geometric
picture of the Higgs mechanism: the $W$ bosons arise from strings
stretching between the branes; their mass is proportional to their
length, which is the separation of the branes.

For type I or heterotic strings, the supersymmetric gauge theories
living in the world-volume of D-branes have 8 super-charges.  In
particular, $N$ coincident type I 5-branes, which corresponds to $N$
point-like $SO(32)$ instantons, have a 6d ${\cal N}=(1,0)$
supersymmetric gauge theory living in their world-volume \cite{Wsmall}
with gauge group $Sp(N)$, 16 matter hypermultiplets in the ${\bf 2N}$
fundamental, and a hypermultiplet in the antisymmetric ${\bf
N(2N-1)}$.  There are $N$ hypermultiplet flat directions associated
with the matter in the antisymmetric representation, which break
$Sp(N) \rightarrow Sp(1)^N$, corresponding to the locations of the $N$
5-branes in the four transverse directions.  Again, the separation of
the branes gives a geometric picture of the Higgs mechanism.

The other basic mechanism for composite gauge invariance in string
theory is via branes which wrap surfaces in the internal
compactification space.  The wrapped branes can yield gauge fields in
the remaining uncompactified dimensions, with masses proportional to
the area of the surfaces which they wrap.  When the surfaces
degenerate, the gauge fields become massless, giving another geometric
picture of un-Higgsing a gauge symmetry.  Much as in \cite{conifold},
the new massless gauge fields give the quantum resolution of the
classical singularity in the moduli space where the internal space
degenerates.

The original example of this was in type IIA string theory
compactified to 6d on a K3 surface \cite{Witvd}.  The local behavior
of $K3$ near a singularity looks like an ALE (asymptotically locally
Euclidean) space, which is a possibly blown-up version of the orbifold
$\IC ^2/\Gamma _G$, where $\Gamma _G$ is a discrete subgroup of
$SU(2)$ which acts on the $S^3$ surrounding a point in $\IC ^2$.  The
subscript $G$ on $\Gamma _G\subset SU(2)$ reflects the famous
correspondence between the discrete subgroups of $SU(2)$, cyclic,
dihedral, tetrahedral, octahedral, and icosahedral, and the ADE groups
$G=A_r$, $D_r$, and $E_{6,7,8}$.  The ALE space has $r=$rank$(G)$
two-cycles and thus $3r$ real blowing up parameters.  

The IIA string on the ALE space has an additional $r$ real parameters,
corresponding to the integral of the NSNS two-form $B$ field over the
$r$ two-cycles.  All together, the $4r$ real scalars combine with
$U(1)^r$ gauge fields, coming from reducing the IIA three-form
potential on the $r$ two-cycles, and fermionic partners, to form a
${\cal N}=(1,1)$ supersymmetric $U(1)^r$ gauge theory in the remaining
six dimensions.  Classically, the $U(1)^r$ gauge theory has no charged
matter.  However, it was argued in \cite{Witvd} that two-branes
wrapped around the $r$ two-cycles yield the $W$ bosons needed to
extend $U(1)^r$ to $G$.  For generic values of the $r$ hypermultiplet
blowing up modes and $B$ fields \cite{AspinB}, $G$ is broken to
$U(1)^r$ by the Higgs mechanism -- the two-cycles are non-degenerate
and thus the $W$ bosons are massive -- but, when the blowing up modes
and $B$ fields vanish, the non-perturbative, composite $G$ gauge
symmetry is restored.  The composite gauge symmetry is necessary for
the duality between IIA on K3 and the heterotic string on $T^4$, where
the enhanced gauge symmetry $G$ is visible classically.

The above analysis can obviously be reduced to fewer dimensions by
additional compactification.  It can also be pushed up to 7d and 8d by
compactification of M theory or F theory, respectively, on the
$K3$.  In all cases, it yields supersymmetric theories with 16
super-charges.  See \cite{NSnotes} for discussion of such theories in
various $d$.

It is also possible to obtain via compactification a variety of gauge
theories with 8 super-charges in 6d, 5d, or 4d (and fewer by further
reduction) via compactification of F theory, M theory, or type IIA,
respectively, on Calabi-Yau threefolds.  In these cases, composite
gauge invariance occurs when a complex surface shrinks to zero size
along a complex curve in the Calabi-Yau.  It is possible to get
essentially any gauge group: $A_r$, $B_r$, $C_r$, $D_r$, $E_{6,7,8}$,
$F_4$ and $G_2$ from various types of singularities
\cite{aspgro,Betal}.

Finally, by combining compactifications with having branes in the
transverse directions, it is possible to obtain a variety of
interesting gauge theories living in the world-volume of the branes.
The dynamics of the world-volume gauge theories ``probes'' that of
space-time.  For example, it was shown in \cite{DouMor} that a type II
brane living at a point on an ALE space in the transverse directions
has a supersymmetric gauge theory living in its world-volume whose
Higgs branch is isomorphic to the ALE space itself, giving a physical
realization of the ``hyper-Kahler quotient'' construction \cite{kron}
of the ALE spaces.  Moving the brane around in the transverse ALE
space corresponds to changing hypermultiplet expectation values on the
Higgs branch.  The general connections between the dynamics of the
world-volume gauge theory on the ``probe'' brane and that of
space-time are:
\begin{description}
\item{1.} Global symmetries on the brane $\leftrightarrow$ gauge
symmetries in space-time.

\item{2.} Moduli space of gauge theory on brane $\leftrightarrow$
moduli space of the brane in spacetime, which is the space-time
itself.  Further, the metric of the corresponding moduli on the brane
is the same (locally) as the space-time metric.

\end{description}
In this way, known results concerning either the world-volume gauge
theories or the space-time string theories which they probe, can lead
to new results about the other.  This has lead to a better
understanding of the field theories which can be obtained on branes
and also a better understanding of string theory and string duality.

\section{``Tensionless strings'' and non-trivial RG fixed points}

It was pointed out in \cite{witcom} that the IIB theory at a $\IC
^2/\Gamma _G$ singularity of K3 is quite different from the IIA story
reviewed in the previous section.  Indeed, the theory in the remaining
six dimensions has ${\cal N}=(2,0)$ supersymmetry, whose multiplets
contain five real scalars and a self-dual tensor field, along with
fermion partners, with no gauge fields at all.  There are
$r=$rank$(G)$ such matter multiplets on $\IC ^2/\Gamma _{G}$.  There
are BPS strings, which are obtained by wrapping the three-brane of the
IIB theory on the $r$ two-cycles of $\IC ^2/\Gamma _G$, whose BPS
tension goes to zero at the point where the $5r$ scalar expectation
values go to zero.  Rather than having massless gauge fields, we get
``tensionless strings.''   

The above ``tensionless string'' theories can also be obtained in the
world-volume of IIA or M theory 5-branes \cite{Sopen}.  The theory in
the world-volume of $N$ parallel 5-branes is the same as that of the
IIB theory at a $\IC ^2/\IZ _N$ singularity.  The 5-branes are
connected by 2-branes, which intersect the 5-branes along strings.
These strings become tensionless when the positions of the branes in
the transverse directions coincide.

There are also 6d ``tensionless string'' theories with ${\cal
N}=(1,0)$ rather than ${\cal N}=(2,0)$ supersymmetry.  The canonical
example is the small $E_8$ instanton \cite{GaHan,SWcom}, which has a
moduli space with two branches which intersect at a point at the
origin, where there is a ``tensionless string'' theory.  Omitting the
free decoupled hypermultiplet corresponding to the position of the
5-brane in the 4 transverse directions, one branch has 29
hypermultiplets and is isomorphic to the moduli space of an $E_8$
instanton.  The other branch has a ${\cal N}=(1,0)$ tensor multiplet
rather than the $29$ hypermultiplets; it is labeled by the expectation
value $\ev{a}\in \IR ^+$ of the real scalar component of a 6d ${\cal
N}=(1,0)$ tensor super-multiplet. In the context of the realization of
\cite{Horwit} of the $E_8\times E_8$ heterotic string as $M$ theory on
$S^1/\IZ _2$, the theory arises in the world-volume of the $M$ theory
5-brane in the presence of the parallel 9-brane\cite{DMW}.  The branch
with the tensor multiplet corresponds to separating the 5-brane and
9-brane a distance $\ev{a }$ in the 11-th dimension of $M$ theory.
The 5-brane and 9-brane are connected by the 2-brane of $M$ theory,
which intersects the 5-brane along a string of tension proportional to
$\ev{a }$.  At the ``tensionless string'' point at the origin, where
the two branches touch, there is a six-dimensional, scale-invariant
theory with a global $E_8$ symmetry.

The above ``tensionless string'' theories can be considered in the
limit $M_P\rightarrow \infty$ where, as usual, gravity effectively
decouples and we are usually left with a low-energy effective quantum
field theory.  It is not a-priori clear what to make of these
``tensionless string'' theories in this limit and there are many
questions one might ask: ``Do they have to be formulated in terms of
the original string theory even though gravity and such can be
decoupled in the $M_P\rightarrow \infty$ limit?  Do they satisfy the
axioms of quantum field theory?  Are they local?  Can they be
described in terms of interacting quantum field theories which have
string-like excitations (like QCD) or is it necessary to formulate
them as some sort of non-critical theory of fundamental strings?  If
it is possible to describe them as an interacting field theory, is
there a manifestly field theoretic UV free Lagrangian which flows to
the same theory in the IR?''

It was argued in \cite{suss,SWcom} that, because these theories occur
in flat six-dimensional Minkiwski space, they do likely satisfy the
axioms of a local quantum field theory, with local energy momentum
tensors and other local fields.  To accommodate the tensionless string
excitations, these quantum field theories would have to be at
interacting, scale-invariant fixed points.  At present, there are no
known 6d UV free Lagrangian field theories which flow to the same RG
fixed points in the IR but, as will be discussed below, it is known
that there are such theories upon reduction to fewer dimensions.  This
definitely shows that, at least upon dimensional reduction, the
``tensionless string'' theories really are interacting local quantum
field theories\footnote{More recently, it has been argued in
\cite{snew} that the theories obtained in the world-volume of type II
or heterotic 5-branes, in the limit $M_P\rightarrow \infty$ with $M_s$
held fixed, are {\it not} local quantum field theories.  They have
string-like excitations and exhibit T-duality upon compactification.
The local quantum field theory discussed above (upon compactification
to 3d) is only the low-energy limit of the new theory associated with
$E_8$ instantons.  At energy $\sim M_s$, string-like excitations
become important.}.

\section{3d ${\cal N}=4$ supersymmetric theories}

In \cite{NStd} certain aspects of 3d ${\cal N}=4$ supersymmetric
(i.e. 8 super-charges, the same number as ${\cal N}=2$ supersymmetry
in 4d) $U(1)$ and $SU(2)$ gauge theories were solved by using
connections with string theory and string duality.  The results were
subsequently rederived purely via a field theory analysis \cite{SWtd},
giving a non-trivial check of string duality.  The argument of
\cite{NStd} is as follows: 3d ${\cal N}=4$ $U(1)$ or $SU(2)$ theories
arise in the world-volume of a type I (or heterotic) 5-brane (or small
instanton) which is compactified on $T^3$, with the 5-brane wrapping
$T^3$ to yield a 2-brane.  The heterotic or type I theory on $T^3$ is
dual to $M$ theory compactified on $K3$, where the 2-brane in the
uncompactified 7d space-time must arise as the fundamental $M$ theory
2-brane at a point in $K3$.  The moduli space of the 3d $U(1)$ or
$SU(2)$ supersymmetric gauge theories must probe, and thus reproduce,
the local geometry of $K3$.

The theory in the uncompactified 7d space-time can have $G=SU(N)$,
$SO(N)$, or $E_{6,7,8}$ gauge symmetry.  In the heterotic description,
the gauge symmetry $G$ arises perturbatively for particular $T^3$
moduli while, in the $M$ theory description, it is composite gauge
invariance associated with compactification on a $K3$ with a $\IC
^2/\Gamma _G$ singularity, where $\Gamma _G$ is the discrete subgroup
of $SU(2)$ corresponding to group $G$.  The 3d theory living in the
world-volume of the 2-brane probe has a moduli space with two branches
which touch at a point at the origin, where there is an unbroken,
global $G$ symmetry.  One branch is isomorphic to the moduli space of
a $G$ instanton.  The other branch is isomorphic to the $\IC ^2/\Gamma
_G$ orbifold space which the theory probes.

The 3d world-volume theory for $G=SU(N)$ is $U(1)$ with $N$ electrons.
The fact that this theory has a Higgs branch, where the electrons get
expectation values, which is isomorphic to the moduli space of an
$SU(N)$ instanton is a classical result which is not modified by
quantum effects (and independent of 3d, depending only on there being
8 super-charges).  This theory also has a Coulomb branch where the
scalars in the photon vector-multiplet get expectation values.
Classically, the Coulomb branch is $\IR ^3\times S^1$, with the $S^1$
the scalar $\gamma $ which is dual to the photon in 3d via
$*dA=d\gamma$.  The above argument of \cite{NStd} shows that, via
quantum effects, the Coulomb branch must become $\IC ^2/\IZ _N$.
Similarly, the 3d world-volume theory for $G=SO(N)$ is $SU(2)$ with
$N$ fundamental quarks.  The fact that this theory has a Higgs branch
which is isomorphic to the moduli space of an $SO(N)$ instanton is a
classical result (which is crucial in \cite{Wsmall}) which is not
modified by quantum effects.  The Coulomb branch, on the other hand,
is modified by quantum effects and must become $\IC ^2/\Gamma
_{SO(N)}$.  The singularities at the origin of these 3d $U(1)$ and
$SU(2)$ theories for $N>1$ implies that they lead to non-trivial RG
fixed points with global $SU(N)$ or $SO(N)$ symmetry respectively
\cite{NStd}.

Finally, the 3d theories associated with $G=E_{6,7,8}$
correspond to reduction of the 6d small $E_8$ instanton ``tensionless
string'' theories discussed earlier.

It was pointed out in \cite{ISmir} that there can be a ``mirror
symmetry'' duality for 3d, $N=4$ supersymmetric theories; this
involves dual descriptions of RG fixed points with the exchanges:
Higgs branch $\leftrightarrow$ Coulomb branch, classical
$\leftrightarrow$ quantum, mass terms $\leftrightarrow$ \FI\ terms,
and manifest global symmetries $\leftrightarrow$ hidden global
symmetries not present in the ultra-violet Lagrangian. Examples
include duals to $U(1)$ with $N$ electrons, $SU(2)$ with $N$ quarks,
and the $E_{6,7,8}$ ``tensionless string'' theories mentioned above.
The dual descriptions of the 3d RG fixed points with global $G=SU(N)$,
$SO(N)$, and $E_{6,7,8}$ symmetry are given, for all $G$, by $\prod
_iU(n_i)$ gauge theories, with each $U(n_i)$ factor associated with a
node of the corresponding $G$ Dynkin diagram and matter given by
bi-fundamentals corresponding to the links of the extended Dynkin
diagram.  For the $G=E_{6,7,8}$ cases, this gives local, UV free,
Lagrangian quantum field theories which flow in the IR to the same RG
fixed points as the small $E_8$ instanton ``tensionless string''
theories! This shows, at least in 3d, that these are {\it local}
quantum field theories at non-trivial RG fixed points\footnote{See,
however, the earlier footnote.}.

The 3d mirror symmetry of \cite{ISmir} was recently related to string
theory dualities.  In \cite{PorZaf}, it was connected with $M$ theory
on $K3\times K3$, with the mirror symmetry related to an exchange of
the (non-compact) K3s.  In \cite{HanWit}, it was connected with
intersecting branes of type IIB theory, with the mirror symmetry
related to the $SL(2,\IZ )$ S-duality of the type IIB theory.  More
recently, it was shown in \cite{HOVdual} that mirror symmetry, in
gauge theories realized geometrically via type II string theory on a
Calabi-Yau threefold and an additional circle, is tantamount to
T-duality.

\section{5d $N=1$ supersymmetric theories}

We now turn to 5d $N=1$ susy theories (which have 8 super-charges, the
same number as the 3d theories discussed in the previous section).
For finite coupling, as discussed at the end of sect. 1, these
theories are not asymptotically free in the UV and are infrared free.
However, it was argued in \cite{NSfd} that for gauge group $SU(2)$ and
$N_f\leq 7$ flavors there are non-trivial RG fixed points at
$g_{cl}=\infty$; in other words, the theories with finite coupling can
be regarded as perturbations of the fixed point by the relevant
operator $\Delta {\cal L}=g_{cl}^{-2}F^2$, which drives the theory to
flow to a free theory in the IR.  Thus the $SU(2)$ theories with
$N_f\leq 7$ really ``exist,'' as they can be defined in the UV at the
scale-invariant fixed point with $m_0\equiv g_{cl}^{-2}\rightarrow
0$. (Note that $g_{cl}^{-2}$ has dimensions of mass in 5d.)  It was
further argued in \cite{NSfd} that the $SO(2N_f)\times U(1)_I$ global
symmetry of $SU(2)$ with $N_f$ flavors for $m_0\neq 0$, where $U(1)_I$
has current $j_I=*(F\wedge F)$, is extended to $E_{N_f+1}$ at $m_0=0$.
In particular, the RG fixed point with $N_f=7$ and $m_0=0$ is that of
the small $E_8$ instanton theory reduced to 5d.

The argument of \cite{NSfd} is via probing type I $\leftrightarrow$
heterotic duality when compactified on $S^1$.  Wrapping the type I
5-brane around the $S^1$ (with $SO(32)$ Wilson lines included) leads
to a 4-brane with a 5d $SU(2)$ world-volume gauge theory with $N_f\leq
16$ flavors.  The full theory is T-dual to type I' on an interval with
sixteen D8 branes and one D4 brane probe in the bulk and orientifold
planes at each of the two ends.  When the probe and $N_f$ of the D8
branes are near an orientifold plane, the theory on the probe is
$SU(2)$ with $N_f$ flavors.  The distance between the probe and the
orientifold plane is $\ev{a }\in \IR ^+$, the real scalar in the
$SU(2)$ gauge multiplet whose expectation value breaks $SU(2)$ to
$U(1)$, giving the Coulomb branch.  The distance between the D8 branes
and the orientifold planes gives the mass of the $SU(2)$ flavors.
When, for example, the $N_f$ flavors are massless, the effective
$U(1)$ gauge coupling on the Coulomb branch is $t(a )F_{\mu \nu }^2$,
where $t(a )$ is {\it exactly} given by supersymmetry, 5d gauge
invariance, and a one-loop calculation to be $t(a )=m_0+(8-N_f)a$; the
$8$ is the contribution of $W$ bosons and the $-N_f$ is the
contribution of quarks running in the loop.  $t(a)$ gives the exact
space-time metric (and thus dilaton), showing that a result of
\cite{PolWit} is exact.

Consider the effective gauge coupling $t(a )=m_0+(8-N_f)a$ as a
function of $\ev{a} \in \IR ^+$.  For $N_f>8$, there is a ``Landau
pole'' on the Coulomb branch for $a \geq m_0/(N_f-8)$; $m_0\neq 0$ is
needed as a UV cutoff.  For $N_f=8$, $t(a )=m_0$ is a constant and it
is impossible to take the limit $m_0\rightarrow 0$ without having the
effective gauge coupling degenerate.  For $N_f<8$ it is possible to
take $m_0\rightarrow 0$ while preserving $t(a )=g_{eff}^{-2}\geq 0$ on
the entire Coulomb branch $\ev{a}\in \IR ^+$.  The theory at $a =0$ is
scale invariant and interacting -- thus it is a non-trivial RG fixed
point.

While this argument of \cite{NSfd} shows that RG fixed points {\it
could} exist for $N_f\leq 7$, one might wonder if they really {\it do}
exist.  Probing type I -- heterotic duality \cite{NSfd} shows that
they do indeed exist, with $E_{N_f+1}$ global symmetry.  In the
context of type 1/heterotic on $S^1$, $m_0$ corresponds to the radius
of the $S^1$, with $m_0=0$ corresponding to $R=R_{critical}$, the
value for the radius where spacetime gauge symmetry is enhanced from
$SO(2N_f)\times U(1)$ to $E_{N_f+1}$.  For the $SO(32)$ heterotic
string on $S^1$, this is visible at the classical level via winding
modes which become massless at $R=R_{critical}$.  For $N_f\leq 7$, at
$m_0=a =0$, there is an $E_{N_f+1}$ gauge symmetry in spacetime and
thus an $E_{N_f+1}$ global symmetry for the theory on the brane.  For
$m_0\neq 0$ the Higgs branch, which intersects the Coulomb branch at
the origin, is isomorphic to the moduli space of an $SO(2N_f)$
instanton; for $m_0\rightarrow 0$ it becomes isomorphic to the moduli
space of an $E_{N_f+1}$ instanton.  The interacting, scale invariant
theories at the origin of the moduli space correspond to the $E_8$
``tensionless string'' theory, reduced to 5d.

Unlike the situation in 3d \cite{ISmir}, reviewed in the previous
section, in 5d there are no known UV free Lagrangian field theories
which flow {\it into} the small $E_8$ instanton ``tensionless string''
RG fixed points in the IR.  Nevertheless, these are local quantum
field theories at RG fixed points \cite{NSfd}.  The Lagrangian $SU(2)$
field theories with $N_F\leq 7$ flavors flow {\it out of} these fixed
points upon perturbation by a relevant operator \footnote{See,
however, the earlier footnote.}.

It is also possible to get 5d $N=1$ supersymmetric $SU(2)$ with $N_f$
flavors as ``composite'' gauge invariance and matter from $M$ theory
on a singular Calabi-Yau threefold.  The physics/geometry dictionary
is that a ruled complex surface, with $N_f$ singularities along the
ruling, collapses to a complex curve of size $\sim m_0$.  $M$ theory
on the CY leads to 5d Chern-Simons interactions $c_{ijk}A^i\wedge
F^j\wedge F^k$, where $c_{ijk}$ is the intersection form for $H_4$
classes.  These terms are related by 5d susy to the effective gauge
coupling.  Since we know $t(a)=m_0+ (8-N_f)$, this means that the
intersection form $c$ must satisfy $c=8-N_f$ for the singularity which
gives $SU(2)$ with $N_f$ flavors; this is indeed the case, giving a
check of $M$ theory and its composite gauge invariance.  Now one can
consider the limit $m_0\rightarrow 0$, which is where the complex
surface $S$ is a ``del Pezzo'' surface which can collapse to a point.
The gauge theory results for $N_f\leq 7$ perfectly matches and extend
the del Pezzo classification \cite{DMNS,DKV}.

In the work \cite{IMS}, we consider general aspects of 5d susy gauge
theories, including an anomaly which renders some theories
inconsistent and others consistent only by including a Wess-Zumino
type Chern-Simons term.  We discuss the general necessary conditions
for non-trivial RG fixed points and find all possible gauge groups and
matter content which satisfy them.  The gauge theory results are
connected, via $M$ theory, to results about Calabi-Yau geometry.  In
particular, the classification of non-trivial RG fixed points related
to gauge theory yields a new classification of the higher codimension
analog of ``del Pezzo'' singularities of CY threefolds.  It is thereby
verified using the geometry that the new 5d RG fixed points do indeed
exist.

5d $N=1$ susy gauge theories have a real adjoint scalar $\Phi$ and
thus there is always a Coulomb branch moduli space with $\ev{\Phi}$ in
the Cartan, breaking $G\rightarrow U(1)^r$.  The Coulomb branch is a
{\it wedge} parameterized by Cartan scalar moduli $\phi ^i$ in $\IR
^r/{\cal W}$, where $\cal W$ is the Weyl group of $G$.  The unbroken
$U(1)^r$ is enhanced to larger $G$ subgroups at the walls of the
Coulomb branch Weyl chamber.  5d $N=1$ susy implies that the low
energy effective theory on the Coulomb branch is given by a
prepotential ${\cal F}(\phi ^i)$, which leads to $U(1)^r$ effective
gauge couplings $t(\phi )_{ij}F^iF^j$ and moduli space metric
$ds^2=t(\phi )_{ij}d\phi ^id\phi ^j$, where $t(\phi )_{ij}=\partial
_i\partial _j {\cal F}(\phi )$ must be continuous.  In addition, there
is a Chern-Simons term $c_{ijk}A^i\wedge F^j\wedge F^k$, where
$c_{ijk}=\partial _i\partial _j\partial _k {\cal F}$.  The
Chern-Simons term is not gauge invariant and thus it is necessary to
have $c_{ijk}\in \IZ$, for all $i,j,k$, in order for the action to be
well-defined mod $2\pi l$, with $l$ integer.  This requires $c_{ijk}$
to be a constant independent of $\phi ^i$; thus, gauge invariance
implies that the prepotential can be at most locally cubic: ${\cal
F}(\phi ^i)=\half t^0_{ij}\phi ^i\phi ^j+\f{1}{6} c_{ijk}\phi ^i\phi
^j\phi ^k$.

The quantization condition $c_{ijk}\in \IZ$ implies that some theories
are only sensible when a classical Chern-Simons term is included.  For
example, $U(1)$ gauge theory with $N_f$ electrons has Chern-Simons
coefficient $c=c_{classical}+c_{quantum}$ where, via a one-loop
calculation \cite{WitMF}, $c_{quantum}=-\half N_f sign(\phi )$; this
result is exact.  Gauge invariance then implies that
$c_{classical}+\half N_f\in \IZ$; thus, for odd $N_f$,
$c_{classical}\neq 0$.  There is a similar situation for non-Abelian
theories. At the classical level, ${\cal F}=\half m_0\Tr \Phi ^2
+\f{1}{6}c_{cl}\Tr \Phi ^3$, where the first term gives the classical
gauge coupling and the second term, which is only possible for groups
with a non-trivial cubic Casimir, i.e. only for $G=SU(N)$ with $N\geq
3$, gives a 5d non-Abelian Chern Simons term.  For groups $G$ with
$\pi _5(G)\neq 0$ there can be a global anomaly \cite{NieSem,IMS}
analogous to the 4d $\pi _4$ anomaly of \cite{Wold}.  The non-trivial
cases are $\pi _5(SU(N))=\IZ$ for $N\geq 3$ and $\pi _5(Sp(N))=\IZ
_2$.  For $SU(N)$, the $\pi _5$ anomaly implies that the theory is
only consistent if $c=c_{cl}+c_{quantum}\in \IZ$; with $N_f$
fundamentals, this requires $c_{cl}+\half N_f\in \IZ$ and thus
$c_{cl}\neq 0$ for $N_f$ odd.  For $Sp(N)$, the $\pi _5$ anomaly
implies that the theory with an odd number of half-hypermultiplets is
inconsistent.

Because the prepotential is at most cubic, it is {\it exactly} given
by a one-loop calculation, which yields the cubic term.  The result,
which is valid for any gauge group $G$ and matter multiplets in
representations $r_f$ with masses $m_f$ is 
\bea {\cal F}=\half m_0h_{ij}\phi ^i\phi ^j+\f{1}{6}c_{cl}d_{ijk}\phi
^i\phi ^j\phi ^k +\f{1}{12}\left(\sum _{\bf R} |{\bf R}\cdot \phi
|^3-\sum _f\sum _{{\bf w}\in {\bf W_f}}|{\bf w}\cdot \phi
+m_f|^3\right),\eea 
where the last two terms are the contributions,
respectively, of the vector and matter multiplets in the loop.
In this expression, $m_0=g_{cl}^{-2}$, $h_{ij}\equiv \Tr (T_iT_j)$,
$d_{ijk}\equiv \half \Tr (T_i\{T_j,T_k\})$, and $\bf R$ are the roots of
$G$ and $\bf W_f$ are the weights of $G$ in representation $r_f$.

It is necessary for the theory to have $ds^2=t(\phi )_{ij}d\phi
^id\phi ^j\geq 0$ and thus $\partial _i\partial _j {\cal F}$ must have
non-negative eigenvalues on the Coulomb branch.  When the $c_{ijk}\phi
^k$ part has negative eigenvalues, the theory is at best sensible in a
subspace of the Coulomb wedge near the origin by taking $m_0>0$.  On
the other hand, when this part is positive, there can be a scale
invariant RG fixed point at the origin with $m_0=0$.  A {\it
necessary} condition for a RG fixed point is thus that, with $m_0=0$,
$\partial _i\partial _j{\cal F}d\phi ^id\phi ^j\geq 0$ throughout the
entire Coulomb wedge.  This is equivalent to the condition that $\cal
F$ be a {\it convex} function over the entire Coulomb wedge --
i.e. for any points $x$ any $y$ in the Coulomb wedge, ${\cal
F}(\lambda x+(1-\lambda )y)\leq \lambda {\cal F}(x)+(1-\lambda ){\cal
F}(y)$ for all $0\leq \lambda \leq 1$.  Using the above general
expression for $\cal F$ it is possible to classify all possible gauge
groups and matter content satisfying this condition.

Note that the gauge contribution to $\cal F$, $\f{1}{12}\sum _{\bf
R}|{\bf R}\cdot \phi |^3$, is purely convex whereas the matter
contribution, $-\f{1}{12}\sum _f\sum _{{\bf w\in W_f}}|{\bf w}\cdot
\phi +m_f|^3$, is purely concave.  Thus there can be a RG fixed point
provided there isn't too much matter.  This is analogous to the
condition of asymptotic freedom in four dimensions.  Some general
comments are: $G$ must be non-Abelian to have a fixed point. Any
non-Abelian theory with no matter can have a RG fixed point.  There
can be no fixed point for theories with matter in representations with
weights $|{\bf W_f}|\geq |{\bf R}|$.  If $\cal F$ is convex, giving
mass to some matter and integrating it out yields a low-energy ${\cal
F}_{low}$ which is even more convex; this is consistent with flowing
between RG fixed points.  There can be no new fixed points associated
with product groups -- matter which couples the groups necessarily
leads to non-convex $\cal F$.

Consider, for example, $G=Sp(N)$ with $n_A$ matter fields in the two
index antisymmetric representation and $N_f$ in the fundamental
representation.  The Coulomb branch wedge is given by $\Phi =diag(a_1,
\dots, a_N, -a_1, \dots -a_N)$/permutations and can thus be chosen to
be $a_1\geq a_2\geq \dots \geq a_N\geq 0$.  The prepotential for
$m_0=g_{cl}^{-2}=0$ is ${\cal F}=\f{1}{6}(\sum
_{i<j}[(a_i-a_j)^3+(a_i+a_j)^3](1-n_A)+\sum _i a_i^3(8-N_f))$.  This
function is convex on the entire Coulomb wedge provided either $n_A=1$
and $N_f\leq 7$ or $n_A=0$ and $N_f\leq 2N+4$.  The RG fixed points
with $n_A=1$ and $N_f\leq 7$ can be shown to exist by probing type I
$\leftrightarrow$ heterotic duality on $S^1$ with $N$ four brane
probes.  This argument shows that the RG fixed points have a
$SP(1)\times E_{N_f+1}$ global symmetry and there is a Higgs branch
which is the moduli space of $N$ $E_{N_f+1}$ instantons.  The
existence of the RG fixed points for $n_A=0$ can be shown via $M$
theory on a singular CY, which reproduces exactly the condition
$N_f\leq 2N+4$.

For $G=SU(N)$, the Coulomb wedge can be taken to be $a_1\geq
a_2\geq \dots \geq a_N$ with $\sum _i a_i=0$. The prepotential (1)
with $m_0=0$ is convex on the entire Coulomb wedge when there are
$N_f$ matter fields in the fundamental representation and $c_{cl}$, 
with $c_{cl}+\half N_f\in
\IZ$ for gauge invariance, which satisfy $N_f+2|c_{cl}|\leq 2N$.
All these can be obtained from $N_f=2N$ and $c_{cl}=0$ by adding
masses of various signs and RG flow.  For $N_f<8$ the prepotential is
also convex with $n_A=1$ two index antisymmetric representation matter
field provided $N_f+2|c_{cl}|\leq 8-N$.  For $SU(4)$ there are also
solutions with $n_A=2$.  

For $G=SO(N)$, we find that $\cal F$ is convex and thus there can be
RG fixed points for $n_V\leq N-4$.  For $N\leq 12$ there can also be
spinors, with $n_S\leq 2^{(12-N)/2}$ for $N$ even and $n_S\leq
2^{(11-N)/2}$ for $N$ odd.  For $G_2$, the necessary condition for RG
fixed points is $N_f\leq 4$; for $F_4$, it is $n_{26}\leq 3$; for
$E_6$, it is $n_{27}\leq 4$; for $E_7$, $n_{56}\leq 3$; for $E_8$,
there can be no added matter fields.  This gives a complete
classification of possible 5d ${\cal N}=1$ RG fixed points related to
gauge theories.  The next question is if they all really exist.  This
will be discussed below.

We now turn to the Calabi-Yau interpretation, with the 5d gauge
theories obtained via $M$ theory on singular CY geometries.  Comparing
with the 5d gauge theory results gives a highly non-trivial check of
the relation in $M$ theory between singularities in CY geometry and
the composite gauge invariance in space-time. In particular, the
geometry of intersecting surfaces on the CY must yield an intersection
form which reproduces the group theory of our general result eq. (1)
for $\cal F$.  The fact that this indeed works in every case is quite
astonishing.

For example, to get $SU(3)$, we need two intersecting, ruled, complex
surfaces, $S_1$ and $S_2$, to collapse to a complex curve of size
$m_0\sim g_{cl}^{-2}$.  The surfaces $S_i$ correspond to Cartan
$U(1)_i$ generated by $(T_i)_{jk}=
\delta _{jk}(\delta _{i,j}-\delta _{i+1,j})$, $i=1,2$.  Their
intersections are $S_1^3=8$, $S_2^3=8-N_f$, $S_1^2S_2+S_1S_2^2=-2$,
where, by an arbitrary choice, the singularities in the ruling
corresponding to the $N_f$ flavors were put in the $S_2$ component.
These relations yield $S_1^2S_2=-1+c'$ and $S_1S_2^2=-1-c'$, where
$c'$ is an undetermined integer.  These relations give 
$$6{\cal F}=\sum _{ijk}c_{ijk}\phi ^i\phi ^j\phi ^k =\sum
_{i<j}(a_i-a_j)^3 +(c'+\half N_f)\sum _ia_i^3-\half N_f\sum
_i|a_i|^3,$$ which exactly agrees with the field theory result (1)!
The term $c'+\half N_f$ is the Chern-Simons term $c_{cl}$ and indeed
satisfies the quantum quantization condition $c+\half N_f\in \IZ$.

Now, if it is mathematically possible to take the size $m_0$ of the
complex curve to zero in the Calabi-Yau construction of enhanced gauge
symmetry $G$ with a given matter content, we'll have a string theory
construction of the corresponding RG fixed point theory.  We have thus
verified in \cite{IMS} that, in (essentially) every case, the above
classification for RG fixed points perfectly matches and extends the
mathematics of higher codimension del-Pezzo collapses.  (We expect it
to also work in the few, more technically difficult, cases which were
not checked in \cite{IMS}.)  Thus, all of the 5d RG fixed points in
our classification do indeed exist.

To conclude, gauge theories have interesting dynamics in various
dimensions.  There are several types of interplay between gauge and
string dynamics.  These interconnections have led to new insights into
field theory, string theory, and mathematics.

\section{Acknowledgments}
I would like to thank D.R. Morrison and especially N. Seiberg for
collaborations on which parts of this talk are based and for many
illuminating discussions.  I would like to thank the organizers for
the interesting conference.  This work was supported in part by NSF
PHY-9513835, the W.M. Keck Foundation, an Alfred Sloan Foundation
Fellowship, and the generosity of Martin and Helen Chooljian.


\end{document}
\vfil
\end